\documentclass[preprint,12pt]{elsarticle}

\usepackage{amssymb}

\journal{Physica A}

\begin{document}

\begin{frontmatter}



\title{Effect of random field disorder on the first order transition in $p$-spin interaction model.}

\author{Sumedha}
\address{National Institute of Science Education and Research, Institute of Physics Campus, Bhubaneswar, Orissa- 751 005, India}

\author{Sushant K Singh}
\address{Variable Energy Cyclotron Centre,1/AF Bidhan Nagar, Kolkata-700064, India} 




\begin{abstract}
We study the random field $p$-spin model with Ising spins on a fully connected graph using the theory of large deviations in this paper. This is a good model to study the effect of quenched random field on systems which have a sharp first order transition in the pure state. For $p=2$, the phase-diagram of the model, for bimodal distribution of the random field, has been well studied and is known to undergo a continuous transition for lower values of the random field ($h$) and a first order transition beyond a threshold, $h_{tp}(\approx 0.439)$. We find the phase diagram of the model, for all $p \ge 2$, with bimodal random field distribution, using large deviation techniques. We also look at the fluctuations in the system by calculating the magnetic susceptibility. For $p=2$, beyond the tricritical point in the regime of first order transition, we find that for $h_{tp}<h<0.447$, magnetic susceptibility increases rapidly (even though it never diverges) as one approaches the transition from the high temperature side. On the other hand, for $0.447<h \le 0.5$, the high temperature behaviour is well described by the Curie-Weiss law. For all $p \ge 2$, we find that for larger magnitudes of the random field ($h>h_o=1/p!$), the system does not show ferromagnetic order even at zero temperature.  We find that the magnetic susceptibility for $p \ge 3$ is discontinuous at the transition point for $h<h_o$. 

\end{abstract}

\begin{keyword}
Random field, p-spin interaction, first order transition, spin-glass


\end{keyword}

\end{frontmatter}




\section{Introduction}
\label{section1}
Magnetic systems in the presence of random field are an important class of disordered systems which have been extensively studied for last many years. These systems are governed by two energies: the exchange interaction between spins which want spins to be correlated and local random field $h$ at each site which would make the spins uncorrelated. Thermal frustrations play a secondary role in random field models, since their disordering effect 
is much weaker \cite{dfisher,binder}. 

One of the very well studied model  to understand the effect of random field is the random 
field Ising model (RFIM), which was introduced by Imry and Ma in 1975 \cite{imryma}. 
The pure model (Ising model) has a continuous transition for all $d \ge 2$. The model has been 
very useful as it has found its physical realizations in many of the condensed matter systems \cite{belanger}. Hence the model has attracted a lot of attention, both theoretically and  experimentally \cite{dfisher,binder,belanger,nettermann,reiger,fytas,sourlas,picco}. The model has many interesting and intriguing properties. For example, at the mean field level the phase-diagram of RFIM is known to change its character as a function of the random field distribution function. For gaussian distribution, the transition stays continuous for all strengths of the random field, but for bimodal distribution it exhibits a tricritical point \cite{aharony}. In three dimensions though the existence of tricritical point has been widely debated \cite{sourlas,picco, binder,hernandez,fytas2}, with recent numerical simulations suggesting continuous transition even in the strong disorder regime\cite{fytas,fytas2}.

Related model is a $p$-spin model, in which $p$-spins are connected by an exchange interaction \cite{gross}. Pure system has a sharp first order transition for $p \ge 3$, which can easily be  studied in the mean field limit. The random field version of this model, for $p \ge 3$, has relatively attracted much less attention in the literature. This model is a good system to 
study the effect of local field randomness on sharp first order transitions \cite{cardy1}. The 
model is also of relevance since it is known to have connections with the structural glasses \cite{kirkpatrick,biroli}.

Effect of quenched local randomness for systems which undergo continuous phase transition in the absence of randomness is well described by the Harris criterion \cite{harris}. This is possible due to the universal behaviour of the correlation length. On the other hand, even though first order transitions are common in nature, effect of random local disorder on first order  transitions is much harder to study theoretically, as the correlation length remains finite and hence the universal behaviour is not expected in these systems \cite{randomfirstorder}.

In this paper we look at the random field $p$-spin model on fully connected graphs using the theory of large deviations. Large deviations lend itself naturally to the study of mean field statistical mechanics models \cite{ellis}. Recently, Lowe et al \cite{lowe}  derived the rate function (to be defined in section 2) for the  magnetisation of random field Ising model on fully connected graphs using a method based on tilted large deviation techniques \cite{tilted}. Their method is much simpler than the earlier approaches \cite{comet} and can easily be generalised to many other random field problems. In this paper, we extend it to solve the random field $p$-spin model for arbitrary $p$. Note that even though calculations are a straightforward generalisation of \cite{lowe}, they are useful as they allow us to study the random $p$-spin model fully and hence allow us to look into the issue of understanding the disorder effects near the first order transitions. Besides obtaining the phase diagram in this paper, we also derive a method to obtain the value of quantities like magnetic susceptibility from the rate function. Magnetic susceptibility is an experimentally measurable quantity, and hence is useful to make direct connection with the experimental observations. We find that in the first order regime of RFIM, magnetic susceptibility is a rapidly increasing function of the temperature as one approaches the critical point, even though it does not diverge at the transition point. For $p \ge 3$,  it is discontinuous at the transition point, with a typical paramagnetic behaviour in the high temperature phase. 

The plan of the paper is as follows: In section \ref{sec2}, we give the details of the model and calculate the rate function. We also derive the expression for magnetic susceptibility in this section. In section \ref{sec3}, we give the phase-diagram obtained by looking at the value of magnetisation, and in section \ref{sec4}, we calculate the magnetic susceptibility for $p=2,3$ and $4$.  We conclude in 
section \ref{sec5}.

\section{Rate function for a random field p-spin model}
\label{sec2}
We consider a Hamiltonian of the form:

\begin{equation}
\label{pspinhamiltonian}
{\cal{H}} = -\frac{1}{p! N^{p-1}} \sum_{i_1,...i_p}^N \prod_{k=1}^{k=p} s_{i_k} -\sum_{i_1}^N h_{i_1} s_{i_1}- H \sum_{i_1}^N s_{i_1}
\end{equation}
where $s_{i_k}$ are Ising spins which take values $\pm 1$, $H$ is the uniform external 
field and $h_i$ is the site dependent random field taken from a distribution. The 
probability of getting a particular configuration $\{s_{i_k}\}$ for a given realization of 
$\{h_i\}$ is given by:

\begin{equation}
\label{eqn_gibbs_measure}
P_{N}(\beta,\{ s_{i_k} \},\{h_i\})=\frac{1}{Z_{N,\beta}^h}\mbox{exp}\left( \frac{\beta}{2N}\sum _{ i_1,....,i_p }^N\prod_{k=1}^{k=p} s_{i_k} +\beta \sum _{i_1=1}^N(h_{i_1}+H) s _{i_1}\right)
\end{equation}
where,
\begin{equation}
Z_{N}(\beta,\{h_i\})=\sum _{\{s_{i_k}\}} \mbox{exp}\left( \frac{\beta}{2N}\sum _{ i_1,....,i_p }^N\prod_{k=1}^{k=p} s_{i_k} +\beta \sum _{i_1=1}^N(h_{i_1}+H) s _{i_1}\right)
\end{equation}
is the partition function of the model for a given realization $\{h_i\}$ of the random field and $\beta$ is the inverse of the temperature. In order to get the free energy, one needs to average the $\log$ of the partition function over the distribution of random fields. This averaging over the $\log$ is a hard problem, and has been done only for a relatively fewer models like the random energy model \cite{derrida}. Another approach is the replica approach where the $<\log Z>$ is approximated by the use of the identity $<Z^n>=1+n <\log Z>+O(n^2)$ \cite{parisi}. One more method which has been found useful to study the mean field spin systems is the large deviation technique. One can prove the existence of large deviation principle with an explicit rate function for the order parameter of the model and then use it to study the phase diagram \cite{ellis}. Recently, for random field Ising model ($p=2$), Lowe et al \cite{lowe} derived the rate function explictly for the probability distribution of magnetisation. For a sequence of random variables $\{s_i\}$, the rate function is defined as

\begin{equation}
\lim_{N \rightarrow \infty} \frac{1}{N} P_N(\{s_i\} \in A)=-I(A),
\end{equation}
where $P_N(\{s_i\} \in A)$ is the probability that the given sequence of random 
variables ($\{s_i\}$), belong to the set $A$. For our model, $A$ can be defined via $x=\frac{1}{N}\sum_{i=1}^{N} s_i$, \emph{i.e.}, all sequences with same value of $x$ will have the same probability, as given by the equation above. 

Note that trying to get the rate function directly, using say Gartner Ellis theorem \cite{touchette}, is quite complicated as $\sum s_i/N$ is not a sum of $N$ independent random variables with respect to the probability measure in Eq. \ref{eqn_gibbs_measure}. In order to calculate the rate function Lowe et al \cite{lowe} developed a method which mainly involved the following steps:

1. Prove large deviation principle (LDP) with respect to a measure $Q$, which is a measure on non-interacting spins and hence is easier to average over the quenched field.

2.  Use titled LDP to calculate the rate function for the probability distribution of magnetisation in the presence of coupling term from the rate function of $Q$.

This method turns out to be very useful especially in the cases where the randomness is not in the the coupling term. Also it gives the full rate function for the probability distribution of the  magnetisation, unlike earlier approaches \cite{salinas,hemmen}. Since the average over the spin variables is done over the $Q$ measure, the method extends straightforwardly to arbitrary $p$ for any discrete spin and for any distribution of the random fields. This is unlike other approaches which for $p>2$, were doable only for the spherical spins\cite{haddad}.

Let $Q_i(s_i)$, for every $i\in \mathbb{N}$, be the probability measure on $\{ 1,-1\}$ induced by the field $h_i+H$. Then 
\begin{eqnarray}
Q_i( 1) &=& \frac{e^{\beta (h_i+H)}}{2\mbox{cosh}(\beta (h_i+H))}, \nonumber\\
Q_i(-1) &=& \frac{e^{-\beta (h_i+H)}}{2\mbox{cosh}(\beta (h_i+H))}.
\end{eqnarray}

\begin{figure}
\centering
\begin{tabular}{cc}
\includegraphics[width=4.4cm]{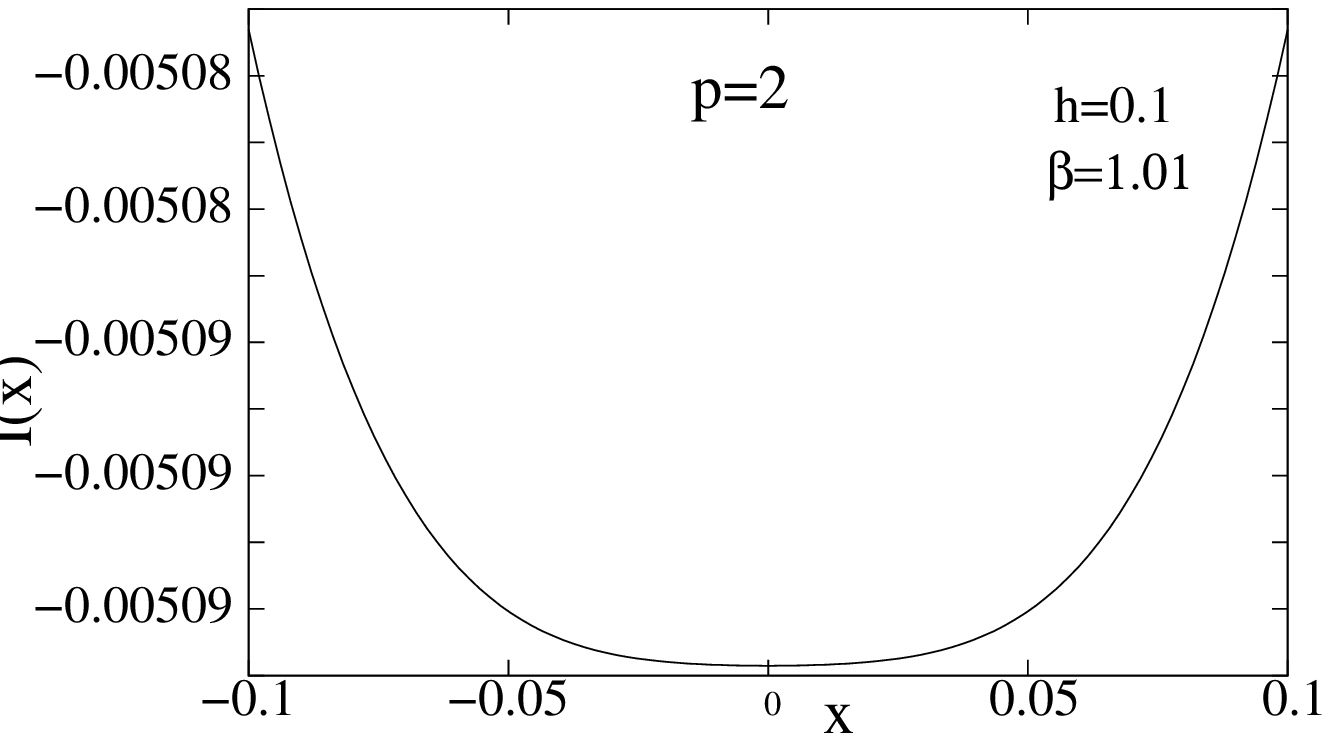} &
\includegraphics[width=4.4cm]{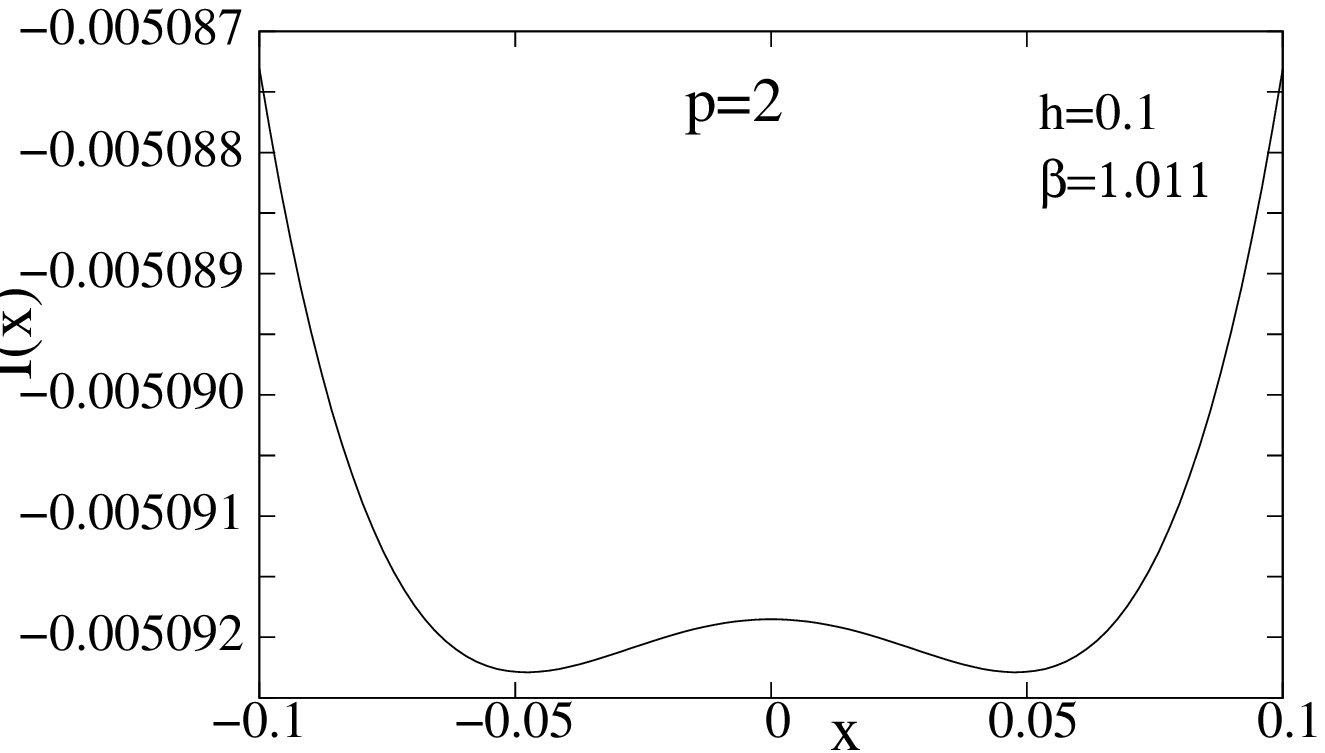}\\ 
\includegraphics[width=4.4cm]{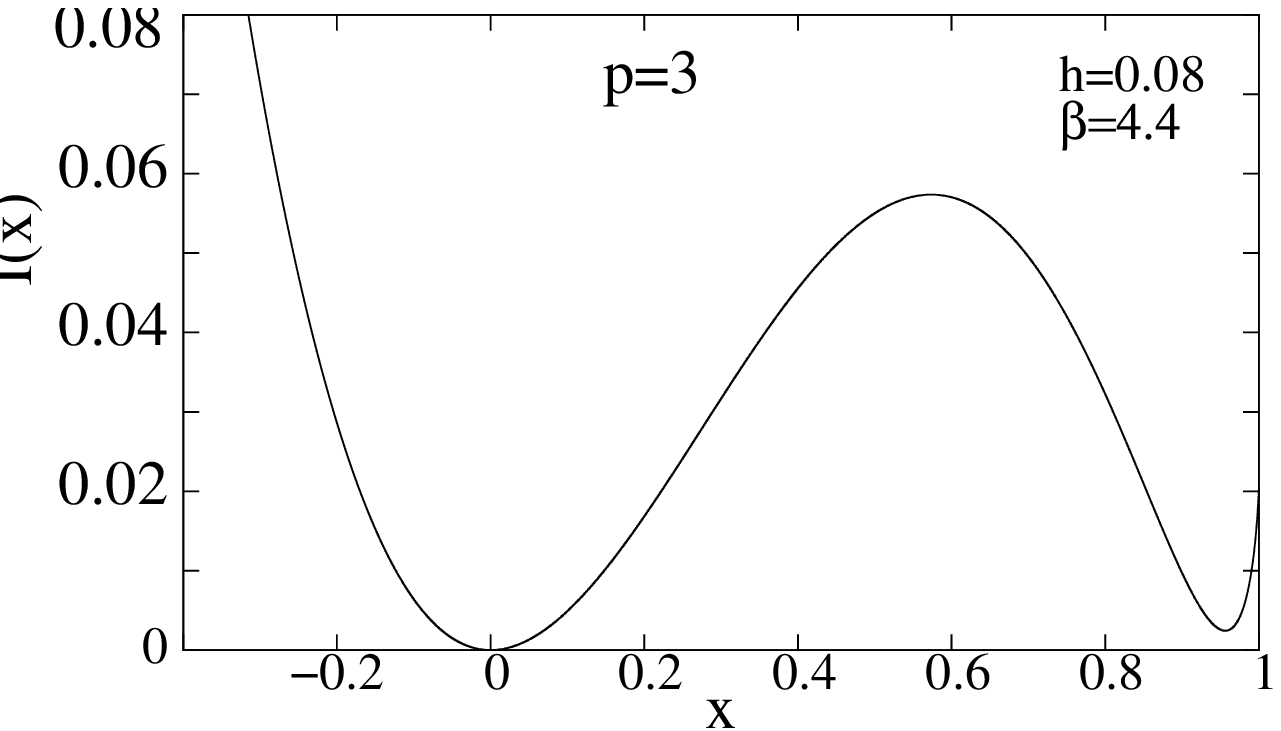} &
\includegraphics[width=4.4cm]{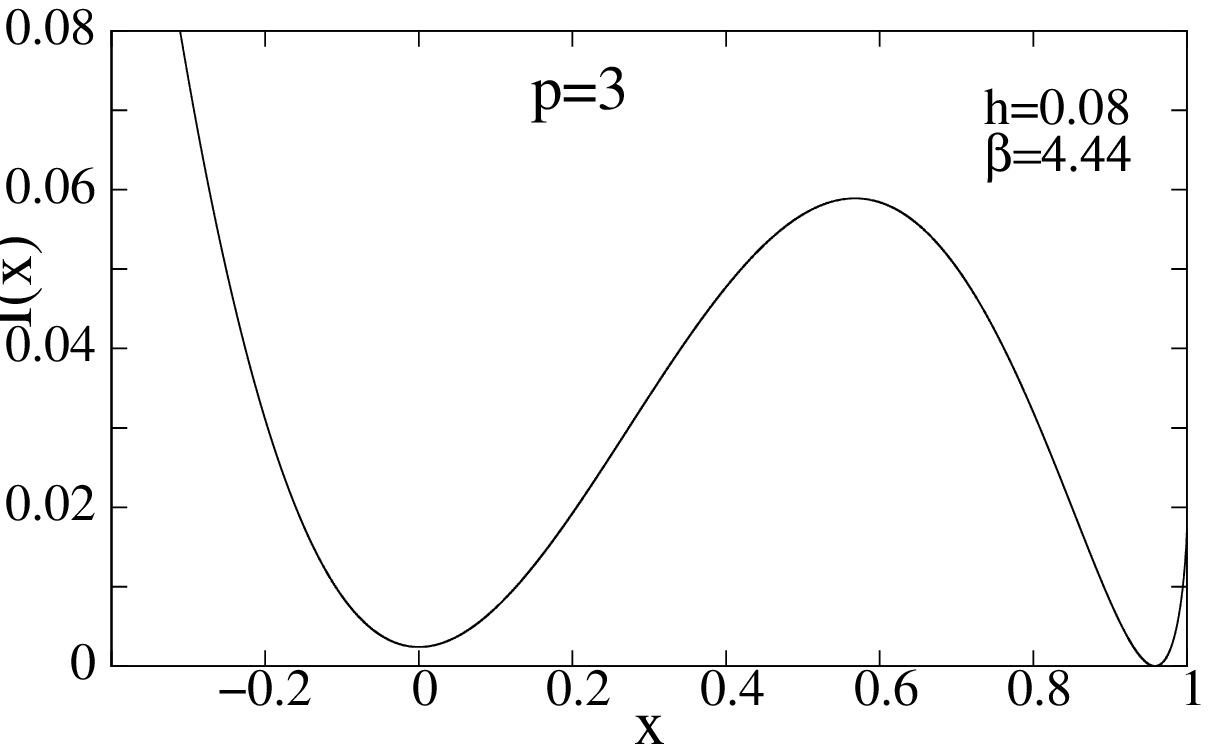}\\ 
\end{tabular}
\caption{The first row shows the rate function ($I_p(x)$) as it changes across a continuous transition for $p=2$. In second row we have plotted the rate function($I_p(x)$) across a first order transition for $p=3$.}
\label{ratef}
\end{figure}

We denote by $Q(\{s_i\})$, the product measure $\otimes _{i=1}^{\infty}Q_i(s_i)$, which is a probability measure on $\{1,-1\}^n$. We first calculate an LDP with respect to $Q(\{s_i\})$. We can employ Gartner Ellis Theorem for the sum, $S_N=\sum_{i=1}^Ns_i$, as it is a sum of independent random variables under the product measure $Q(\{s_i\})$. Gartner-Ellis Theorem yields that $S_N/N$ satisfies LDP with respect to $Q(\{s_i\})$, with a rate function $R(x)=\sup _{y\in \mathbb{R}}\left\{ xy-\Lambda(y)\right\}$, where 
\begin{eqnarray}
 \Lambda (x)&=& \lim _{N \rightarrow \infty}\frac{1}{N}\mbox{ln }\mathbb{E}e^{NxS_N/N}\nonumber\\
 &=& \lim _{N \rightarrow \infty}\frac{1}{N}\sum _{i=1}^{N}\mbox{ln }\left( \frac{e^{x+\beta h_i+\beta H}+e^{-x-\beta h_i-\beta H}}{2\mbox{cosh}(\beta (h_i+H))}
\right) \nonumber\\
&=&\lim _{N \rightarrow \infty}\left( f_N(x)-f_N(0)\right).
\end{eqnarray}
Now because of the Strong Law of Large numbers, we have
\begin{equation}
\label{fn_converge}
\lim _{N\rightarrow \infty}f_N(x)\rightarrow f(x),
\end{equation}
where $f(x)$ is the expectation value with respect to the distribution of the random fields. For the bimodal distribution of the random field $p(h_i)$, given by 
\begin{equation}
\label{eqn_prob_dist_2spin}
p(h_i)=\frac{1}{2}\left[ \delta (h_i-h)+\delta (h_i+h)\right],
\end{equation}
 the function $f(x)$ comes out to be 
\begin{eqnarray}
f(x)=\frac{1}{2}\mbox{ln}[\mbox{cosh}(x+\beta h+\beta H)\mbox{cosh}(x-\beta h+\beta H)].
\end{eqnarray}
Hence we get,
\begin{equation}
R(x)=\sup _{y\in \mathbb{R}}\left\{ xy-\Lambda(y)\right\}=f^*(x)+f(0)
\end{equation}
where $f^*(x)$ is the Legendre-Fenchel transform of $f(x)$,
\begin{equation}
f^*(x)=\sup _{y\in \mathbb{R}}\left\{ xy -f(y)\right\}.
\end{equation}
Now $P_{N,\beta}^h(x)$, the probability that the sum $S_N/N=x$ for a given $\beta$, $h$ and $H$, can be writtem in terms of $Q(\{s_i\})$ as follows:
\begin{eqnarray}
P_{N,\beta}^h(x) &=& \frac{\exp(\beta N x^p/p!) Q(\{s_i\}:\sum s_i=Nx)}{\sum_{Nx=-N}^N \exp(\beta N x^p/p!) Q(\{s_i\}:\sum s_i=Nx)}\\&=& \frac{\exp(\beta N F(x)) Q(\{s_i\}:\sum s_i=Nx)}{\sum_{Nx=-N}^N \exp(\beta N F(x)/2) Q(\{s_i\}:\sum s_i=Nx)}
 \end{eqnarray}
where,
\begin{eqnarray}
F(x)=\left\{\begin{array}{lr}
\beta x^p/p!& \mbox{ if } |x|\le 1\\
\beta /p! & \mbox{ otherwise }.
\end{array}\right.
\end{eqnarray}
The tilted LDP \cite{tilted} then yields that $S_N/N$ satisfies a LDP with respect to $P_{N,\beta}^h(x)$ with the rate function given by
\begin{eqnarray}
\label{eqn_p2_random_rate}
I_p(x)&=& R(x)-F(x)-\inf _{y \in \mathbb{R}}\left\{ R(y)-F(y)\right\}\nonumber \\
&=& f^*(x)+f(0)-F(x)-\inf _{y \in \mathbb{R}}\left\{ f^*(y)+f(0)-F(y)\right\}\nonumber \\
&=& f^*(x)-\frac{\beta x^p}{p!}-\inf _{y \in \mathbb{R}}\left\{f^*(y)-\frac{\beta y^p}{p!}\right\}\nonumber \\
&=& \sup _{y\in \mathbb{R}}\left\{ xy-f(y)\right\}-\frac{\beta}{p!}x^p-\inf _{y\in \mathbb{R}}\left\{ \sup _{z\in \mathbb{R}}\left\{ yz-f(z)\right\}-\frac{\beta}{p!}y^p\right\}.
\end{eqnarray}
Let
\begin{eqnarray}
W(y)=xy -f(y)&=& xy-\frac{1}{2}\mbox{log}[\mbox{cosh}(2(y+\beta H))+\mbox{cosh}(2\beta h)].
\end{eqnarray}
The function $W(y)$ attains maximum at
\begin{equation}
y=\frac{1}{2}\mbox{sinh}^{-1}\left\{ \frac{x}{1-x^2}(b+\sqrt{1+x^2a^2})\right\}-\beta H,
\end{equation}
where $a=\mbox{sinh}(2\beta h)$ and $b=\mbox{cosh}(2\beta h)$.
Hence the rate function in Eq.(\ref{eqn_p2_random_rate}) becomes
\begin{eqnarray}
\label{eqn_p2_random_rate_final}
 I_p(x) &=&\mbox{log }2-\frac{\beta}{p!}x^p-\beta x H+\frac{x}{2}\mbox{Arcsinh}\left( \frac{x}{1-x^2}(b+\sqrt{1+x^2a^2})\right)\nonumber\\&+&\frac{1}{2}\mbox{log}\left( \frac{1-x^2}{2}\right) -\frac{1}{2}\mbox{log}\left(b+\sqrt{1+x^2a^2}\right)-\inf _{y \in \mathbb{R}}G_1(y)
\end{eqnarray}
where, $$G_1(y)=\sup _{z\in \mathbb{R}}\left\{ yz-f(y)\right\}-\frac{\beta}{p!}y^p.$$
At a given value of $\beta,h$ and $H$, the value of magnetisation $m$, which is the average value of $\frac{1}{N} \sum_{i=1}^N s_i$, averaged over all spin configurations, will be given by that value of $x$ at which $I_p(x)$ is minimum. Hence magnetisation is given by that solution of the equation $I_p'(x)=0$, which minimises $I_p(x)$. Rate function defined by Eq. \ref{eqn_p2_random_rate_final}, has one local minimum at $x=0$ in the paramagnetic phase and develops multiple minima in the ferromagnetic phase. This  rate function actually looks very similar to the Landau free energy functional (see Fig. \ref{ratef}). This is not a priori obvious due to the 
presence of random field. Expanding Eq. \ref{eqn_p2_random_rate_final} in powers of $x$ for 
$h=0$ and $H=0$, gives the following expressions for $p=2$, $3$ and $4$ respectively:

\begin{equation}
I_2(x) = (1-\beta) x^2+\frac{1}{6} x^4 +O(x^6),
\end{equation}

\begin{equation}
I_3(x)=x^2-\frac{\beta}{3} x^3+\frac{1}{6} x^4+O(x^6),
\end{equation}

and 

\begin{equation}
I_4(x)= x^2+\frac{(2-\beta)}{12} x^4+\frac{1}{15} x^6+O(x^8).
\end{equation}

As expected, the expansion of $I_2(x)$ in the absence of randomness gives the Landau free energy functional of the Curie Weiss model. Interestingly, for $p \ge 3$, even in the pure case, the $\beta$ dependence is no longer with the lowest power of $x$. Since the expansion is not really useful to get the behaviour near the first order transition (for which we need to keep terms till all orders), we will not use this method to study the phase diagram of the problem in the presence of randomness. The full rate function contains much more information than just the value of magnetisation. We can easily use it to extract the magnetic susceptibility of the system. Magnetic susceptibility is the response of the system to external magnetic field ($\chi=\frac{\partial x}{\partial H}|_{H \rightarrow 0}$).

If $I_p'(x)$ gives us an equation of the form:
\begin{equation}
\label{pminima_eqn}
\frac{\beta}{(p-1)!}x^{p-1}+\beta H-\mathcal{F}(x,\beta,h)=0,
\end{equation}
where $\mathcal{F}$ contains those terms of $I_p'(x)=0$, that are independent of $H$ 
and $p$. Differentiating the above expression with respect to the external field $H$ and then taking the 
limit $H \rightarrow 0$, we get
\begin{equation}
\frac{\beta}{(p-2)!}\beta x^{p-2} \chi +\beta =\frac{\partial \mathcal{F}}{\partial x}\chi
\end{equation}
which gives,
\begin{equation}
\chi =\frac{\beta}{\frac{\partial \mathcal{F}}{\partial x}-\frac{\beta}{(p-2)!} x^{p-2}} .
\label{sus}
\end{equation}
We will use this expression to study the fluctuations in random field $p$-spin model in Section \ref{sec4}.

\section{Phase-diagram}
\label{sec3}
For $p=2$, the above rate function has already been studied \cite{lowe} and we will only briefly mention the results here. For $p=2$ one can actually study a simpler
function, using the fact that the global extremal
points of a function $F(x)-\frac{1}{2}||x||^2$ coincides with the global maximum points of
$\frac{1}{2}||x||^2-F^*(x)$, where $F^*$ is the Legendre Fenchel transform of $F(x)$ \cite{touchette1}. Hence the minima of $I_p(x)$ will coincide with the minima of a function $G(x)$, which is equal to:
\begin{equation}
G(x)=\frac{\beta}{2}x^2-\frac{1}{2}\mbox{log}(\mbox{cosh}[\beta (x+h)]\mbox{cosh}[\beta (x-h)]).
\end{equation}
Interestingly this matches with the expression of free energy for random field Ising model by Salinas and Wreszinski  \cite{salinas}. Studying the minima of this function, one finds that the system undergoes a continuous transition as long as $1-\beta \mbox{sech}^2 (\beta h)<0$ and a first order transition beyond it. At equality, the system has a tricritical point at $h_{tp}= \frac{2}{3} \mbox{cosh}^{-1} \sqrt{3/2}(\approx 0.439)$ and $\beta=3/2$. 

\begin{figure}
\centering
\includegraphics[width=6 cm]{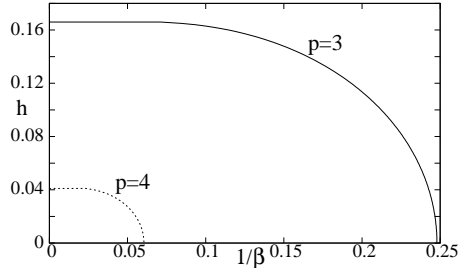} 
\caption{Phase diagram for $p=3$(solid line) and $p=4$(dotted line). Below the line system is in ferromagnetic phase and above it, it is in paramagnetic phase. For $h>h_o \approx 0.167$ for $p=3$, and for $h>h_o \approx 0.041$ for $p=4$, the model has no ferromagnetic phase, i.e, it has zero magnetisation for all values of $\beta$ for $h > h_o$}
\label{figpd}
\end{figure}

For $p \ge 3$, we found that the system either has first order transition to the ferromagnetic state or no transition at all. As shown in Fig. \ref{ratef},  the rate function develops new minima discontinuously. We find that for all $p \ge 2$, the system fails to order into a 
ferromagnetic state, even at zero temperature, for $h>1/p!$. For example, 
for $p=3$, beyond $h_o \approx 0.167$, the magnetisation in the system is always 
zero. Phase diagram for $p=3$ and $p=4$ are plotted in Fig. \ref{figpd}.

\section{Magnetic susceptibility}
\label{sec4}
We want to now look at the fluctuations in the magetisation, as given by the magnetic 
susceptibility of the system. Note that $\chi(x,\beta,h)$ is a function of magnetisation itself, and hence one needs to get the value of $x$ for which $I_p(x)$ is minimum, and then substitute it in the equation of $\chi(x,h,\beta)$ to get the physically acceptable value of magnetic susceptibility at a given $h$ and $\beta$.

\subsection{p=2}
For $p=2$, in the paramagnetic phase, $\chi(0,\beta,h)$ has a particularly simple expression:

\begin{equation}
\label{chizero}
\chi(0,\beta,h) = \frac{2 \beta}{1-2 \beta+ \mbox{cosh}(2 \beta h)}.
\end{equation}
For $h=0$, the above equation gives a divergence in $\chi$ at $\beta_c=1$. But for 
$h \neq 0$, it gives two values of $\beta$ where $\chi(0,\beta,h)$ diverges for all $h<0.447$. We find that for $h<h_{tp}$, the lower value of $\beta$ is the physically relevant one, as beyond that value of $\beta$, magnetisation is no longer zero. For $h_{tp}<h<0.447$, the $\beta$ at which the transition occurs is less than the values of $\beta$ that one obtains by equating $\chi(0,\beta,h)^{-1}$ to zero.

For $p=2$, near the continuous transition (for $h<h_{tp} \approx0.439$), we can study the magnetic susceptibility by keeping terms only till $x^3$ in Eq.(\ref{pminima_eqn}) by expanding $\mathcal{F}(x,\beta, h)$ around $x=0$. We get
\begin{equation}
-\beta H +\frac{x}{2} (1-2 \beta+\mbox{cosh}(2 \beta h)) -\frac{x^3}{3}[\mbox{cosh}^4(\beta h) \{-2+\mbox{cosh}(2 \beta h)\}]=0.
\end{equation}
Differentiating with respect to $H$ and taking the limit $H \rightarrow 0$, we get:
\begin{equation}
\label{p2chi_approx}
\chi(0,\beta,h) = \frac{2\beta}{[1-2 \beta +\mbox{cosh}( 2 \beta h)]-2 x^2 \mbox{cosh}^4(\beta h) [\mbox{cosh} (2 \beta h) -2]}.
\end{equation}
Putting $h=0$ in the above equation, one gets $\chi \sim \frac{1}{1-\beta}$ for $\beta <1$ and
$\chi \sim \frac{1}{2 (\beta-1)}$ for $\beta >1$, as expected. Hence, in the absence of random
field, we reproduce the results of standard mean-field theory. 

For small values of $h$, we can further expand the cosh term and in that regime, the magnetisation is given by:
\begin{eqnarray}
m &=& 0 \hspace{1.4in} \mbox{, for} \hspace{0.1in} \beta <\beta_c \nonumber\\
&=& \sqrt{3 (-1+\beta-\beta^2 h^2)}  \hspace{0.055in} \mbox{, for} \hspace{0.1in} \beta>\beta_c,
\end{eqnarray}
where $\beta_c= \frac{1\pm \sqrt{1-4 h^2}}{2 h^2}$.

Substituting in the expression of $\chi$, we get, for $\beta<\beta_c$
\begin{equation}
\chi(0,b,h) \sim \frac{\beta}{1-\beta+\beta^2 h^2},
\end{equation}
and for $\beta>\beta_c$
\begin{equation}
\chi(m,b,h) \sim \frac{\beta}{2(1-\beta+\beta^2 h^2)}.
\end{equation}

\begin{figure}[!htb]
\centering
\begin{tabular}{cc}
\includegraphics[width=4.4cm]{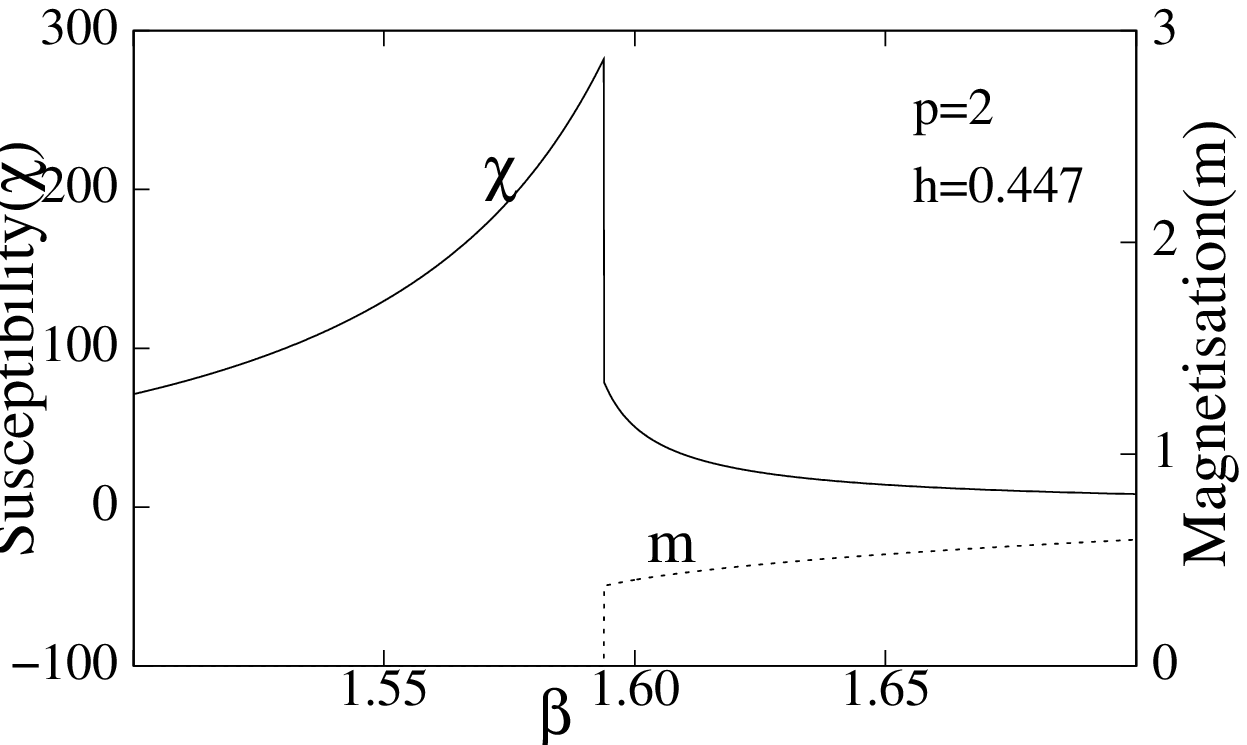} &
\includegraphics[width=4.4cm]{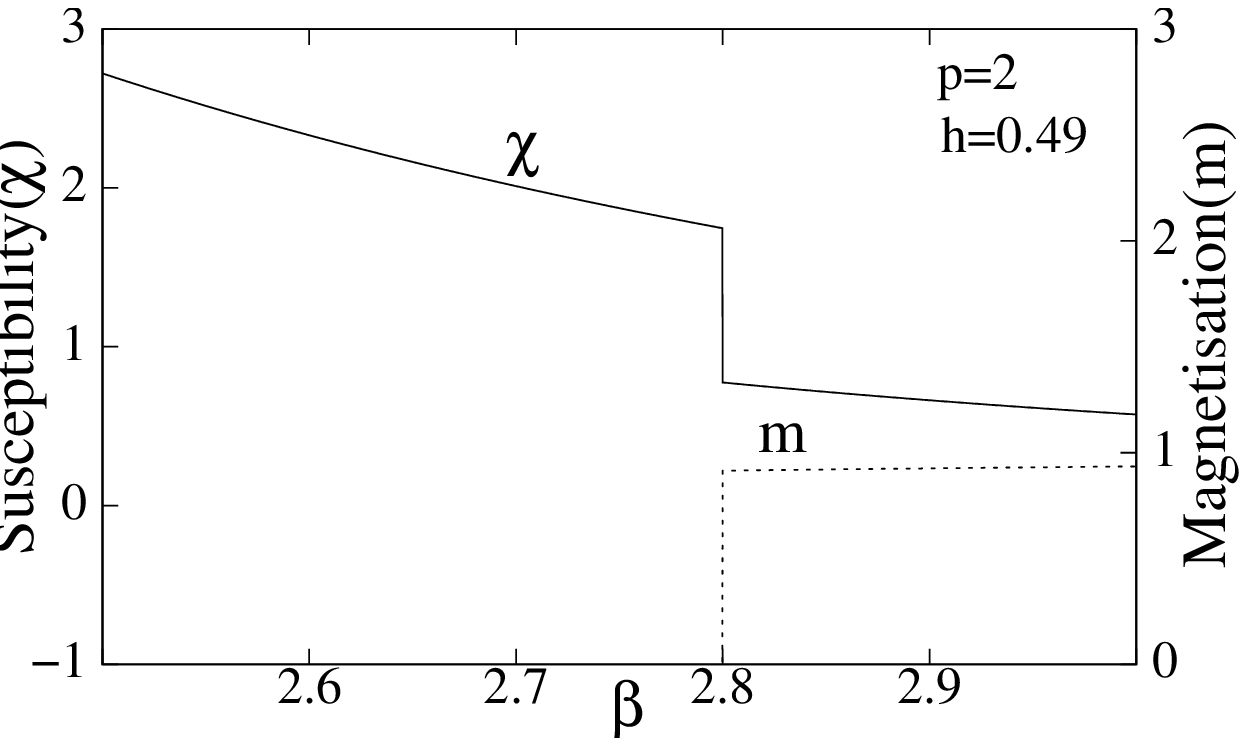}\\ 
\end{tabular}
\caption{Plot of $\chi$ as a function of $\beta$ for $h>h_{tp}$}
\label{fig3}
\end{figure}
For $h>h_{tp}(\approx 0.439)$, the system undergoes a first order transition.  For $h>h_{tp}$, since the system has a first order transition we have to solve Eq. \ref{pminima_eqn} and Eq. \ref{sus} numerically to obtain the magnetic susceptibility. We have plotted magnetic susceptibility and magnetisation for a few representative values of $h$ in Fig 3. We find that for $h_{tp} < h \le 0.447$, $\chi$ is rapidly increasing as one approaches the transition point from the high temperature phase. On the other hand, for $h>0.447$, $\chi$ peaks at a value of $\beta<\beta_c$, and then decreases as it approaches $\beta_c$. This difference in behaviour of magnetic susceptibility arises due to the fact that for $h_{tp}<h< 0.447$, even though the system has a first order transition, the function $\chi(0,\beta,h)$ still shows divergence at two values of $\beta$. In this regime, none of them is relevant as none of them correspond to the transition point and occur well inside the ordered regime. But due to their presence, for $h_{tp}<h<0.447$, $\chi$ still increases rapidly as one approaches $\beta_c$.

\subsection{$p > 2$}

\begin{figure}[!htb]
\centering
\begin{tabular}{cc}
\includegraphics[width=4cm]{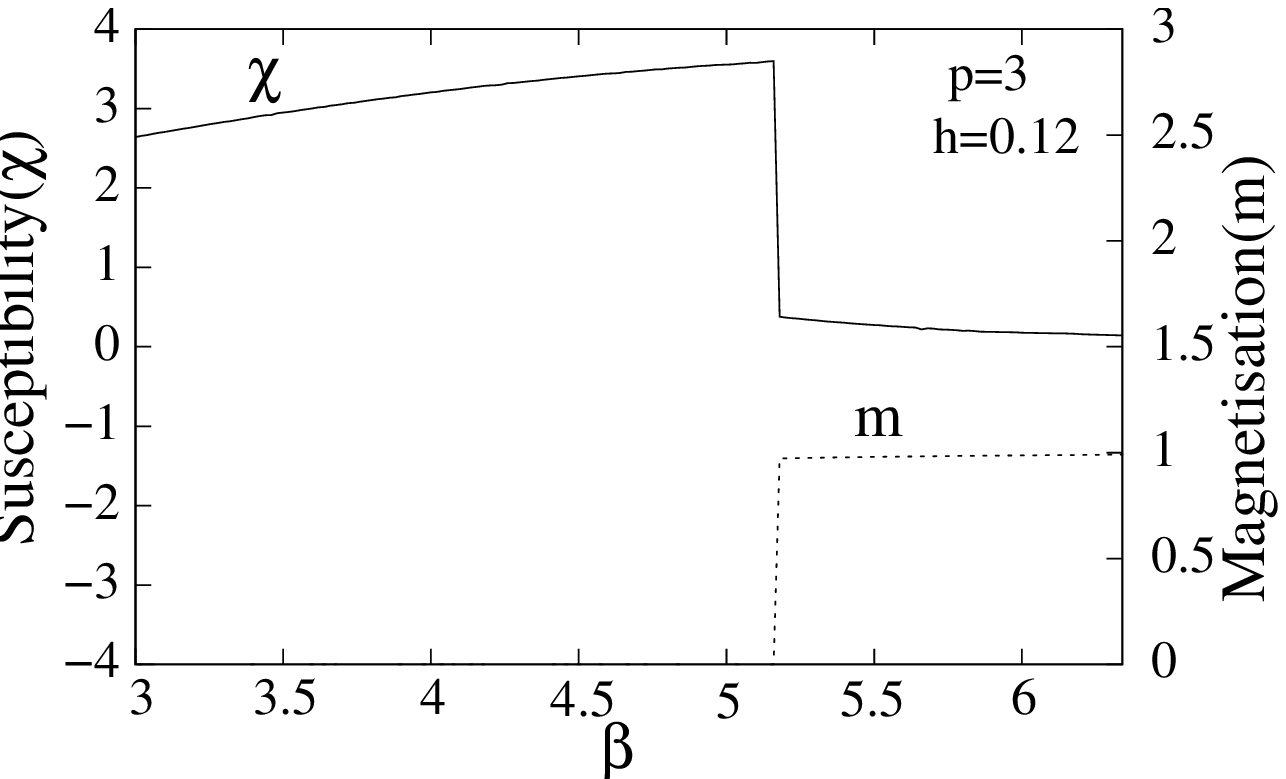} &
\includegraphics[width=4cm]{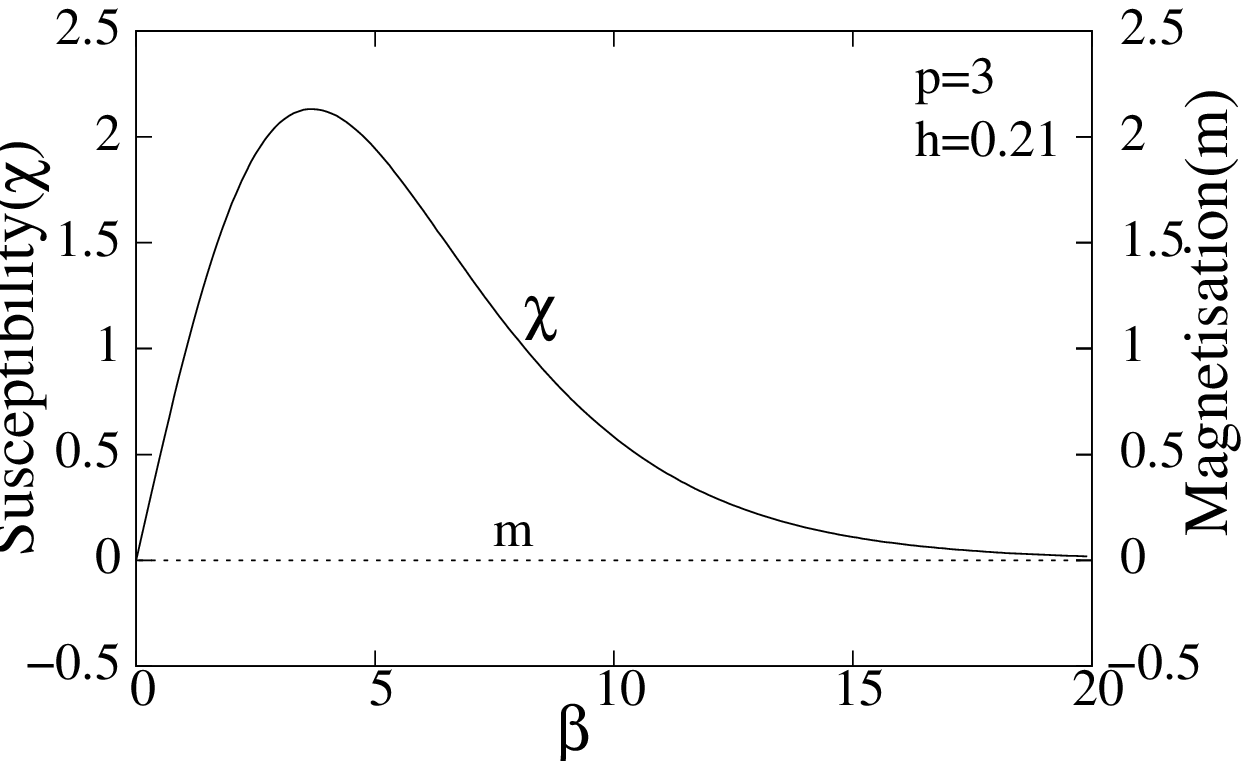}\\ 
\end{tabular}
\caption{$m$ and $\chi$ for $p=3$ for $h<0.167$ and for $h>0.167$.}
\label{fig4}
\end{figure}

\begin{figure}[!htb]
\centering
\begin{tabular}{cc}
\includegraphics[width=4.4cm]{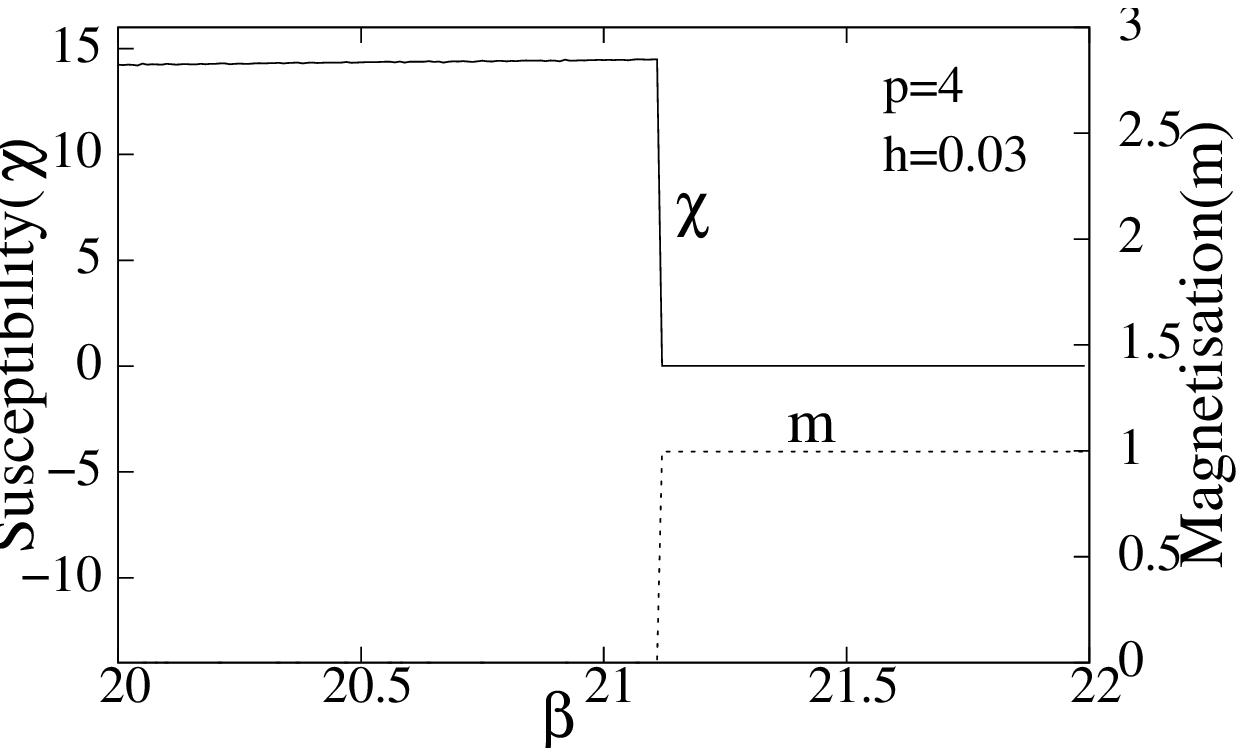} &
\includegraphics[width=4.4cm]{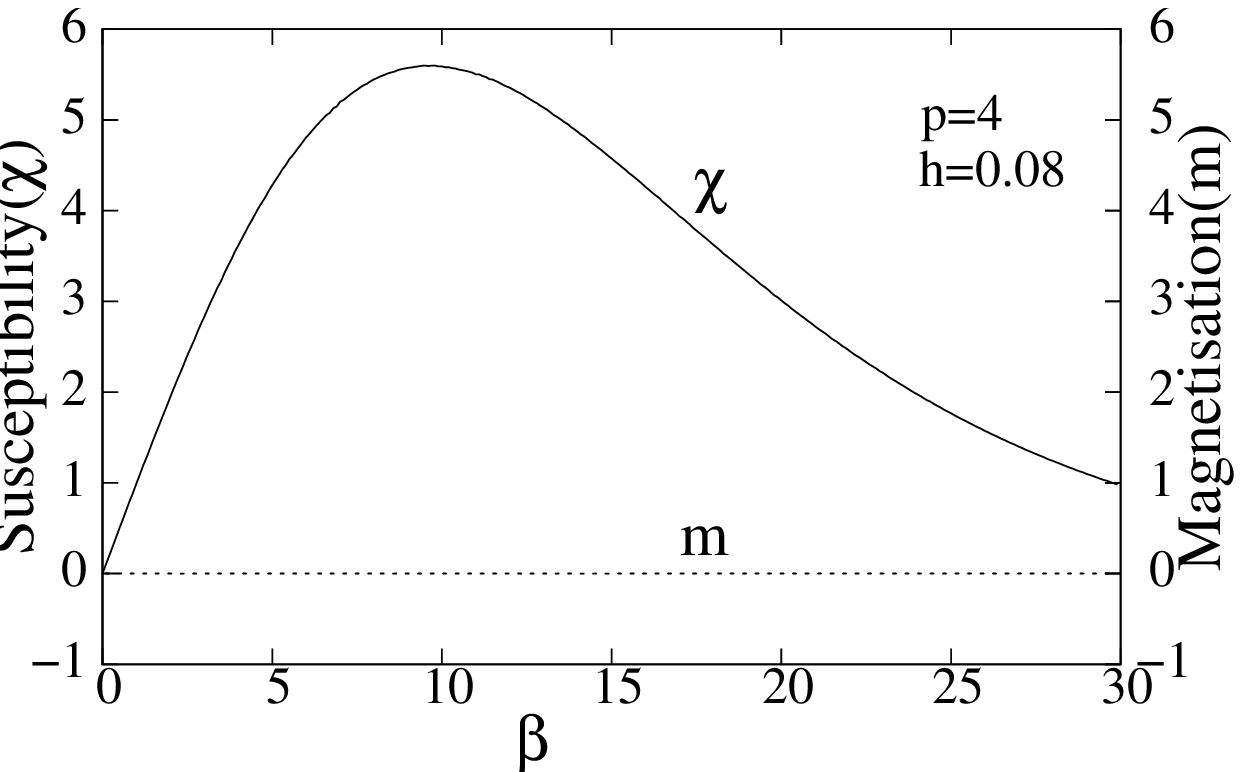}\\ 
\end{tabular}
\caption{$m$ and $\chi$ for $p=4$ for $h<0.041$ and for $h>0.041$.}
\label{fig5}
\end{figure}

For $p=3$ and $p=4$, we numerically evaluate the expression in Eq. \ref{sus} and find that the 
$\chi(0,\beta,h)$ is a convergent function for all values of $\beta$ and $h$. Magnetic susceptibility shows a discontinuity, as expected, at the transition point. In Figs. 4 and 5,  we have plotted magnetic susceptibility  and magnetisation for a few representative values of $h$ for $p=3$, and $p=4$, respectively. For low $\beta$, below the transition, $\chi \propto \beta$ as expected for a paramagnet. Interestingly for $h>1/p!$, even though the system 
does not order ferromagnetically, magnetic susceptibility shows a non monotonic behaviour 
as a function of $\beta$ as shown in Fig \ref{fig4}. Similar behaviour of magnetic 
susceptibility is also seen for $p=4$ beyond $h>0.041$ (Fig. \ref{fig5}).

\section{Conclusion}
\label{sec5}
The nature of phase transition of $p$-spin random field model is qualitatively different 
for $p=2$ and for $p>2$. For $p=3$, the random field model has been solved for spherical 
spins earlier \cite{haddad}. We solve it here for Ising spins, and find that for all $p \ge 2$, for high enough strength of the random field, Ising spins do not order into a ferromagnetic phase, even at zero temperature. Magnetic susceptibility for these strengths of fields also shows a non monotonic behaviour, peaking at a finite $\beta$. The method can be used starightforwardly to study other random distributions, though for $p \ge 3$, we do not expect any qualitative difference in the behaviour of the model for different distributions.

We also notice that the nature of fluctuations near the transition point is different for the first order transition when $p=2$, from that of $p \ge 3$. For $p=2$ there exists a range of $h$ for 
which even though the system undergoes a first order transition, $\chi$ increases rapidly as one approaches the transition. It would be interesting to see if the similar behaviour continues in finite dimensions. In three dimension, RFIM is known to have strong finite size effects which makes it difficult to acertain the nature of transition for the higher strengths of the random field \cite{fytas,sourlas}. Besides that, Eq. \ref{eqn_p2_random_rate_final} can also give us some indications for writing down the correct field theory for the model, which in turn could be useful in understanding the behaviour of randomness near a first order phase transition \cite{cardy1,cardy2}. It should also be possible to extend the method to study the effect of random field in other models like  Potts and Blume-Capel model \cite{sumedha}.

\section{Acknowledgements}
Sushant K Singh would like to acknowledge the local hospitality and support received at NISER  Bhubaneswar, while part of this work was done.

\end{document}